%% file: PaperArXiv.tex
\documentclass[%
 reprint,
 longbibliography,
superscriptaddress,
showpacs,preprintnumbers,
 amsmath,amssymb,
 aps,
pra,
]{revtex4-1}

\usepackage{graphicx}
\usepackage{dcolumn}
\usepackage{bm}

\usepackage{amsmath} %
\usepackage{amsfonts} %
\usepackage{amssymb} %

\usepackage{hyperref}

\usepackage[notbib]{tocbibind}

\begin{document}


\title{Distributional fixed-point equations for island nucleation in one dimension: The inverse problem} 

\author{Hrvojka Krcelic}
  \email{hrvojka.krcelic@strath.ac.uk} 
\affiliation{Department of Chemical and Process Engineering, University of Strathclyde, Glasgow G1 1XJ, UK}  
  
\author{Michael Grinfeld}
\email{m.grinfeld@strath.ac.uk}
\affiliation{ Department of Mathematics and Statistics, University of Strathclyde, Glasgow G1 1XH, UK }%
  
\author{Paul Mulheran}
 \email{paul.mulheran@strath.ac.uk}
\affiliation{Department of Chemical and Process Engineering, University of Strathclyde, Glasgow G1 1XJ, UK}

\date{\today}

\begin{abstract}

The self-consistency of the distributional fixed-point equation
(DFPE) approach to understanding the statistical properties of
island nucleation and growth during submonolayer deposition is 
explored. We perform kinetic Monte Carlo simulations, in which  
point islands nucleate on a one-dimensional lattice during 
submonolyer deposition with critical island size $i$, and examine 
the evolution of the inter-island gaps as they are fragmented by 
new island nucleation. The DFPE couples the fragmentation 
probability distribution within the gaps to the consequent gap 
size distribution (GSD), and we find a good fit between the DFPE 
solutions and the observed GSDs for $i = 0, 1, 2, 3$. Furthermore, 
we develop numerical methods to address the inverse problem, 
namely the problem of obtaining the gap fragmentation probability 
from the observed GSD, and again find good self-consistency in the 
approach. This has consequences for its application to 
experimental situations where only the GSD is observed, and where 
the growth rules embodied in the fragmentation process must be 
deduced.

\end{abstract}

\maketitle


\section{\label{sec:level1}INTRODUCTION}

Island nucleation and growth during submonolayer deposition is a 
topic of continuing research, with ongoing development of 
theoretical models to describe the scaling properties of the island 
sizes and spatial distribution \cite{kolokvijum}. Over the past two 
decades, the focus has tended to move from the problem of obtaining 
the correct form of the island size distribution (ISD) to finding 
the capture zone distribution (CZD) 
\cite{re24,jpd1,jpd2,Mich,EP2017}. 
An island's capture zone is defined as the region on the substrate 
closer to that island than to any other. It represents the growth 
rate of the island, since the deposited monomers that are inside the 
capture zone are most likely to be trapped by the parent island; the 
CZD is therefore a consequence of the spatial arrangement of the 
islands.

One common theoretical approach utilises rate equations, with 
capture numbers reflecting the capture zones 
\cite{re24,jpd1,jpd2,re21,re23}, 
although often this requires some empirically determined 
parameter(s). An alternative approach, of the type we adopt here, 
treats the process using fragmentation equations. An island 
nucleation means the creation of a new capture zone and the size 
reduction of the zones that were previously occupying that region of 
the substrate; so the parent capture zones are fragmented to create 
the daughter capture zone 
\cite{Mich,EP2017,BM,KenMonDen,asim,tokFr,EP2011}.

Aside from these analytical models, when looking only at the 
functional form of a simulated or experimentally obtained CZD in the 
scaling regime, the semi-empirical Gamma distribution function has 
been used frequently as a fitting model, both in two
\cite{11,EP2017,13,14,lit33} and in one dimension \cite{15}. 
Similarly, Einstein and Pimpinelli proposed a Generalized Wigner 
surmise \cite{gws} as a model function for the CZD, relating it back 
to a fragmentation process. An advantage is that functional form can 
be used to deduce the island nucleation mechanism from a measured 
CZD \cite{gws5}, through the critical island size $i$ for nucleation 
(the critical size is defined as the size above which an island will 
not dissociate into monomers). This distribution has also been 
applied to some experimental 
data \cite{lit33}\cite{lit34}\cite{lit35}, however there are also  
some controversies about the 
validity of this model \cite{shi}\cite{liter4}\cite{Mich}.

In the present work, we adopt and explore a nucleation
model on a one-dimensional substrate (modelling nucleation along a 
step edge, for example) in which island nucleation is seen as a 
fragmentation of an inter - island gap. The evolution of the gap 
size distribution (GSD) and CZD is tracked by considering the parent 
gaps (capture zones) that were fragmented by a new island's 
nucleation.

In previous work it was proposed that the GSD and CZD can be 
modelled with distributional fixed-point equations (DFPEs) 
on one-dimensional 
substrates \cite{Ken}. The model equation for the GSD consists 
entirely of physical, easily measurable quantities so in this paper 
we focus solely on the gaps.

The DFPE for the GSD reads:
\begin{align} \label{1}
x \triangleq a(1+x),
\end{align}
where $x$ is a gap size scaled to the average at a given time 
(coverage) and $a$ is a position in the gap where a new island 
nucleates, scaled to the size of the gap ($a \in [0,1]$).

Equation \eqref{1} then says that the distribution of scaled gap 
sizes $x$ is equal 
to distribution of gap sizes that are created when a larger, parent 
gap of size $x+y$ (and, by employing the mean field assumption, 
we've set the scaled size $y = 1$) fragmented into two gaps, of 
proportions $a$ and $1-a$.
DFPE \eqref{1} has an integral equation form:
\begin{align} \label{2}
\phi (x) = \int_0 ^{\min (x,1)} \phi \left( \frac{x}{a} - 1 \right) 
\frac{f(a)}{a} da,
\end{align}
where $\phi(x)$ is the probability distribution function for scaled gap sizes $x$ 
 and $f(a)$ is the probability of breaking a gap into proportions 
$a$ and $(1-a)$. A version of the DFPE \eqref{1} that doesn't 
involve a mean field approximation can also be found in Ref. 
\cite{Ken}, however its corresponding integral equation does not 
offer the possibility of calculating $f(a)$ from a known $\phi(x)$ 
so we will not use it in the present work.
 
Blackman and Mulheran \cite{BM} proposed an analytical form 
for $f(a)$:
\begin{align}\label{3}
f(a) = \frac{(2 \alpha +1) !}{(\alpha !)^2} a^\alpha (1- a)^\alpha .
\end{align}

Here $\alpha$ reflects the mechanism of island nucleation: for 
$\alpha=i$ nucleation is deposition driven and for $\alpha=i+1$ it 
is diffusion driven \cite{KenMonDen}. In a diffusion driven 
nucleation, an island is formed by coming together of $(i+1)$ 
diffusing monomers; in a deposition driven nucleation a smaller, 
unstable cluster is increased to the required $i+1$ size through a 
monomer deposition next to or on top of it. We assume that the real 
(experimental) nucleation process can be modelled as a combination 
of these two idealized cases.  

Eqn. \eqref{3} is derived from the monomer density solutions 
$n_1 (x)$ of a long time (steady - state; $dn_1(x)/dt \simeq 0$) 
diffusion equation with constant monolayer deposition rate 
within a gap \cite{BM}\cite{Mich}. The nucleation probability is 
then assumed to be $\sim n_1 (x)^\alpha$ which gives Equation 
\eqref{3}; therefore it is only valid after the system has had time 
to reach steady state conditions in which monomer density and, by 
extension, $f(a)$ within a gap are time independent. Since the 
deposition rate is constant, provided there is no desorption we 
have coverage $\theta = Ft$. Then Eqn. \eqref{3} is only valid in a 
scaling regime where the GSD, scaled to the average size, as well 
as $f(a)$ are independent of $\theta$.

In this paper, we look further at the applicability of the DFPE 
approach for island nucleation and growth in one dimension. We are particularly 
interested in whether the DFPE provides a self-consistent approach 
to understanding the statistics of gaps.  Two questions are 
addressed:
\\
1. We can measure $f(a)$ during a kinetic Monte Carlo (kMC) simulation; how does the 
measured form compare to that of Eqn. \eqref{3}, and how does the 
solution of Eqn. \eqref{2}, using the observed $f(a)$, compare to 
the kMC GSD?
\\
2. Can we invert the argument of Eqn. \eqref{2}: can we find $f(a)$ 
from a given GSD, and if so how does this recovered $f(a)$ compare 
to that observed in the kMC?

\section{\label{sec:level12} KINETIC MONTE CARLO SIMULATION}

We use a standard kMC simulation model, where 
monomers are deposited onto a one-dimensional lattice with a 
constant monolayer deposition rate $F$, and are free to diffuse by 
nearest-neighbour hopping with diffusion constant $D$. Immobile 
point islands nucleate according to values of the critical island 
size $i$ and subsequently grow by capturing either diffusing or 
deposited monomers. Island nucleation and growth are irreversible 
and re-evaporation of monomers from the surface is forbidden.

We start with an initially empty lattice with $N=10^6$ sites and 
diffusion to deposition ratio $R=D/F=10^7$. We allow monomers to hop 
on average 20 times before the next deposition event 
($R=0.5 \cdot 20 \cdot N$). In total we deposit $n=10^6$ monomers to 
get $\theta = 100 \%$ coverage (not all of the $n$ monomers will get 
incorporated into islands, typically at the end of a simulation 
there is up to a hundred free monomers in the $i=1$ case, and more 
for higher $i$).

At each diffusion step a monomer is selected at random and moved by 
a unit length on the lattice, in a random direction. 
If it arrives to a position adjacent to another monomer or cluster 
of monomers, and the resulting number of monomers is larger than 
$i$, they will be fixed in a single lattice site and the newly 
nucleated island's size and position will be recorded. Islands 
capture monomers that diffuse to adjacent sites and monomers that 
are deposited on top or on an adjacent position. Increments in 
island sizes are recorded while the islands are kept as single 
points on the lattice; this way 
the islands don't coalesce for large coverage, which allows us to 
collect a lot of data 
while the system has still got a long way to go before the scaling 
breaks down \cite{16}. 

In the $i=0$ case we set the probability that a monomer will stick 
to the site onto which it hopped or was deposited, 
to be $p=10^{-7}$.

\begin{figure}[htbp!]
\includegraphics[width=8cm]{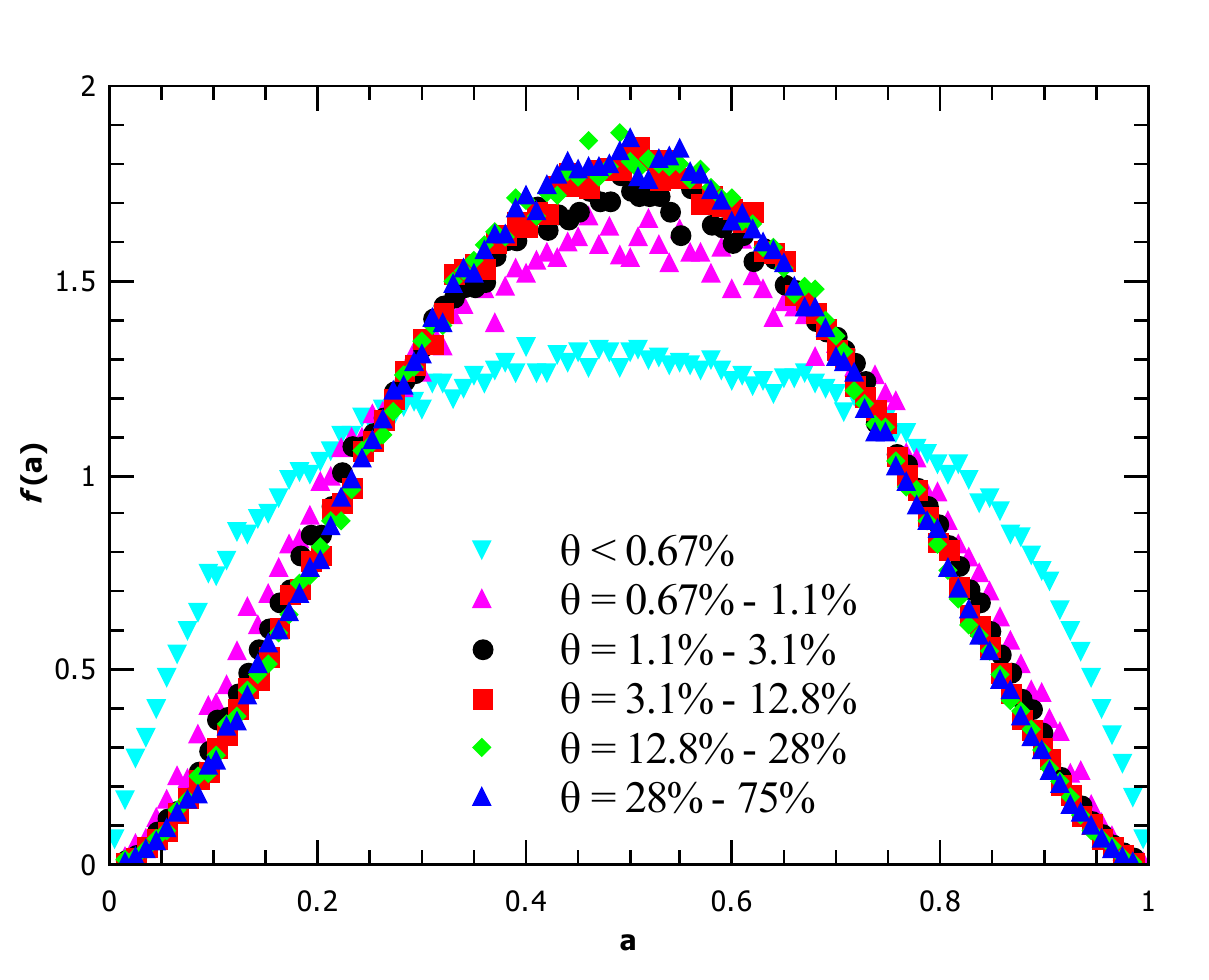}
\centering
\caption{$f_{kMC}$ for various coverage intervals, 
from $\theta = 0 \%$ to $\theta = 75\%$, for $i=1$. Between 
$\theta = 1 \%$ and $3 \%$ the system has reached steady - state 
conditions so subsequent $f_{kMC}$ curves overlap. }
\label{fig:P}
\end{figure}

To get the GSDs ($\phi_{kMC}(x)$) we used outputs at coverage 
$\theta = 100 \%$ and averaged the data over 100 runs.
Every time a new island nucleated, we recorded its position within 
the gap and used that data to create $f_{kMC}(a)$ (as a histogram). 
Since we want to understand scaling properties when the steady state 
conditions have been achieved, we need to find the coverage at which 
$f_{kMC}$ stabilizes. It was previously shown in 
Ref. \cite{KenMonDen} that the monomer density $n_1$ behaves in a 
manner that would yield Eqn. \eqref{3} for small gaps, but not large 
ones. Those findings  corresponds to large coverages (where we 
expect to find mainly  smaller gaps) versus small coverages (large gaps). Figure \ref{fig:P} shows $f_{kMC}$ for $i=1$, reaching 
steady - state condition above the coverage of 
approximately $\theta = 1 \%$.

\section{\label{sec:level13} DFPE METHODOLOGY}

For a given $f(a)$, Eqn. \eqref{2} is solved iteratively for 
$\phi(x)$. Following the procedure described in Ref. \cite{Ken}, we 
perform numerical integration on a mesh of $500$ equally spaced 
points for $x \in [0,5]$. With an initial guess of a rectangular 
$\phi$, we iterate Eqn. \eqref{2} until the solution stabilizes to 
at least its third decimal place.

To solve the inverse problem of obtaining $f$ from a given $\phi$, 
we use two different strategies.

\subsection{Tikhonov regularisation for the Inverse Problem}

Eqn. \eqref{2} belongs to the well-known class of Fredholm integral 
equations of the first kind, $\phi(x) = \int k(x,a) f(a) da$,
which are ill-posed. We also have an additional complication of 
having the left hand side $\phi(x)$ appearing in the kernel function 
$k(x,a)$. This means that any noise 
in the input data will propagate in the kernel. Hence 
this is not a 
standard inverse problem and, to the best of our knowledge, there is 
no established
way of solving this particular type of problem. We proceed to treat 
Eqn. \eqref{2} as we would treat a standard Fredholm equation.

One of the most common ways to deal with ill-posed equations is the 
Tikhonov regularisation procedure, in which a regularisation term 
is added onto the 
original equation. This is a standard method found in many textbooks 
(see for example Ref. \cite{Hansen}); we will describe it briefly.

\begin{figure}[htbp!]
\includegraphics[width=8cm]{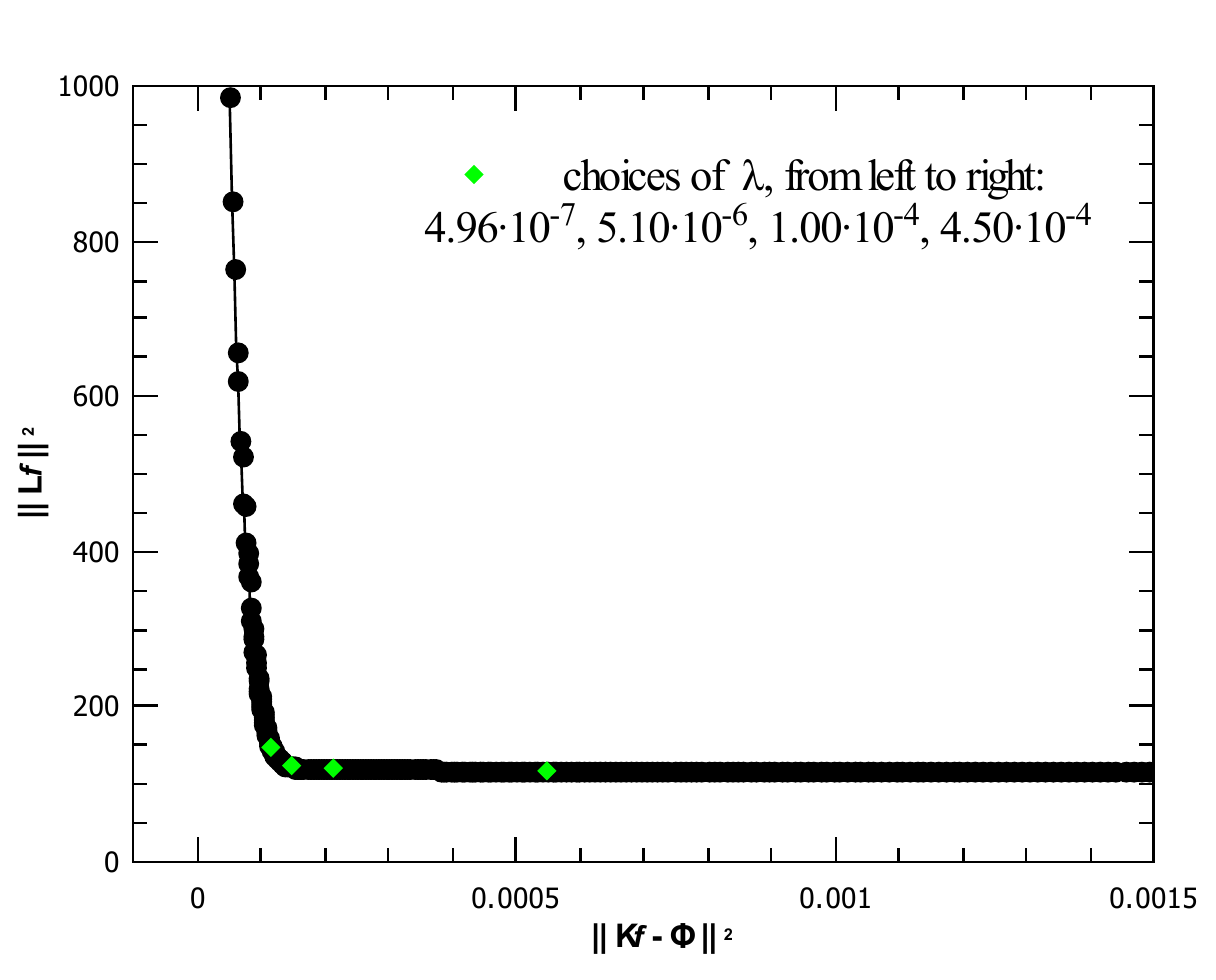}
\centering
\caption{L-curve for inverting Eqn. \eqref{2} with $\alpha = 1$ and 
$f$ defined by Eqn. \eqref{3}. Four chosen values of $\lambda$ are 
marked with green diamond symbols; corresponding solutions 
$P_\lambda$ and their 
integrals $\phi _\lambda$ are shown on Figure \ref{fig:2}.}
\label{fig:1}
\end{figure}

The problem of finding $f$ that satisfies the matrix equation 
$Kf=\phi$ (the discretized form of Eqn. \eqref{2}, where the 
operator $K$ stands for the kernel function and the integral 
operator) can be treated as a minimization problem: 
$\min _{f} \lbrace \lVert Kf - \phi \rVert ^2 _2 \rbrace$.
By adding a regularisation term this problem 
is substituted with: 
$\min _{f} \lbrace \lVert Kf - \phi \rVert ^2 _2 +
 \lambda \lVert L f \rVert ^2 _2 \rbrace$. 
Here, $f$ is the sought solution, $L$ is the regularisation 
operator, usually chosen to be the identity operator or a 
differential operator and 
the regularisation parameter $\lambda > 0$ 
controls how much weight is given to minimization of 
$\lVert Kf - \phi \rVert _2 ^2$ relative to the minimization of the 
added term $\lVert Lf \rVert _2 ^2$.
Solving the inverse problem then includes choosing the appropriate 
operator $L$ and optimizing for $\lambda$.
For a particular value of $\lambda$, the matrix equation to be 
solved for $f$ is:
\begin{align} \label{1.1}
(K^T K + \lambda L^T L ) f_\lambda = K^T \phi
\end{align}
where $K^T$ is the transpose of the matrix operator $K$.
It is straightforward to solve Eqn. \eqref{1.1} numerically.

If $K$ is ill conditioned, with an ill determined rank, the addition 
of the regularisation operator $L$ has a function of making 
Eqn. \eqref{1.1} well posed; then Eqn. \eqref{1.1} will have a 
unique solution $f_{\lambda}$ for all $\lambda$.

The procedure then involves solving Eqn. \eqref{1.1} while varying 
$\lambda$ to find an optimal value of $\lambda$ which stabilizes the 
solution without over - smoothing it. Good values of $\lambda$ are 
typically taken to be within the corner 
of the $\lVert L f \rVert _2 ^2$ vs. $\lVert Kf-\phi \rVert _2 ^2$ 
plot; the so-called L-curve.

We tested the identity and the second derivative operator as 
candidates for the regularisation operator $L$ and, despite the fact 
that second derivative should be the first choice for damping 
oscillatory behaviour in unstable solutions, we found that we get 
better results when using the identity operator. In our 
calculations, we used routines from Ref. \cite{biblija} (see chapter 
therein on Linear Regularisation Methods).

\begin{figure}[htbp!]
\includegraphics[width=8cm]{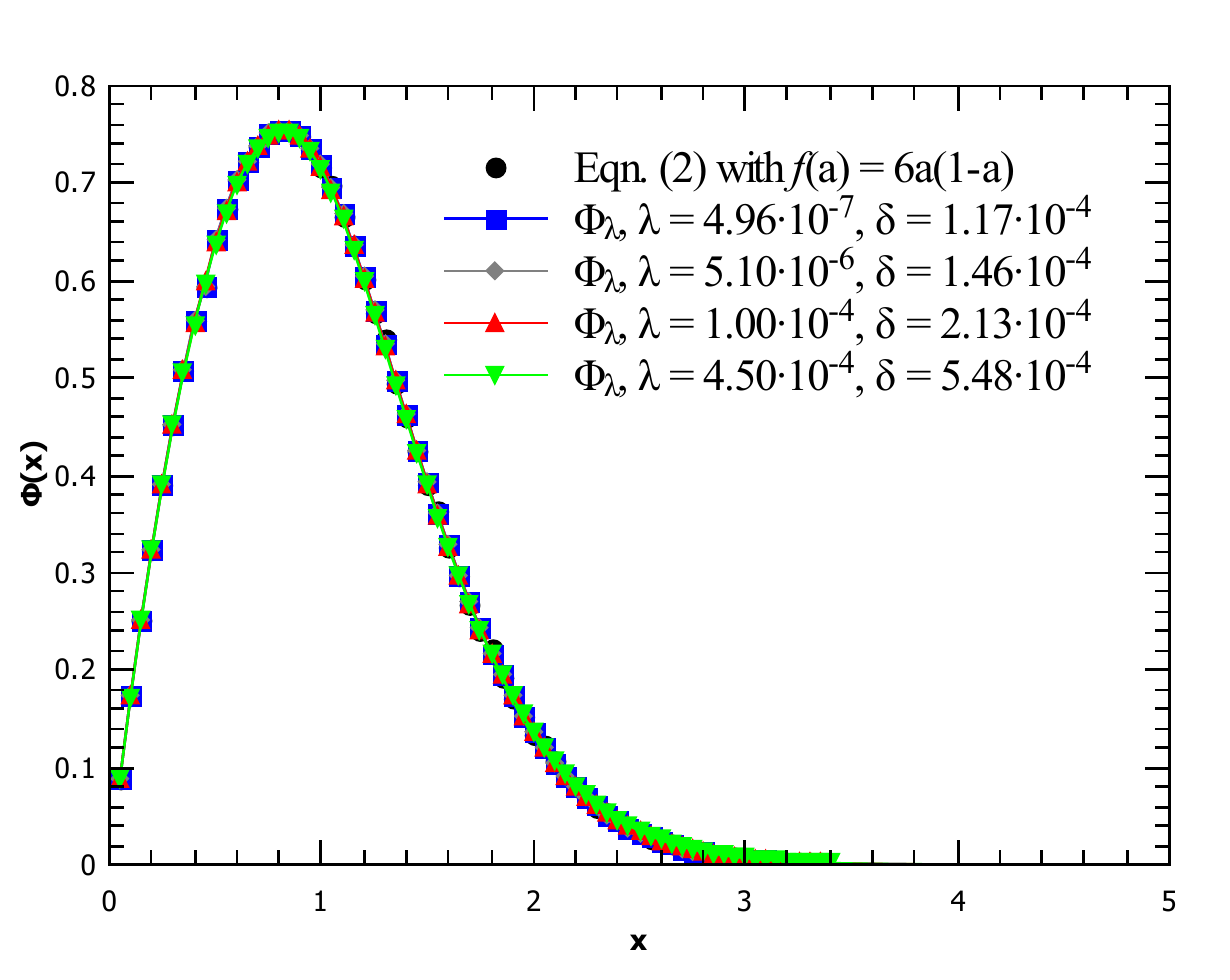}
\centering
\includegraphics[width=8cm]{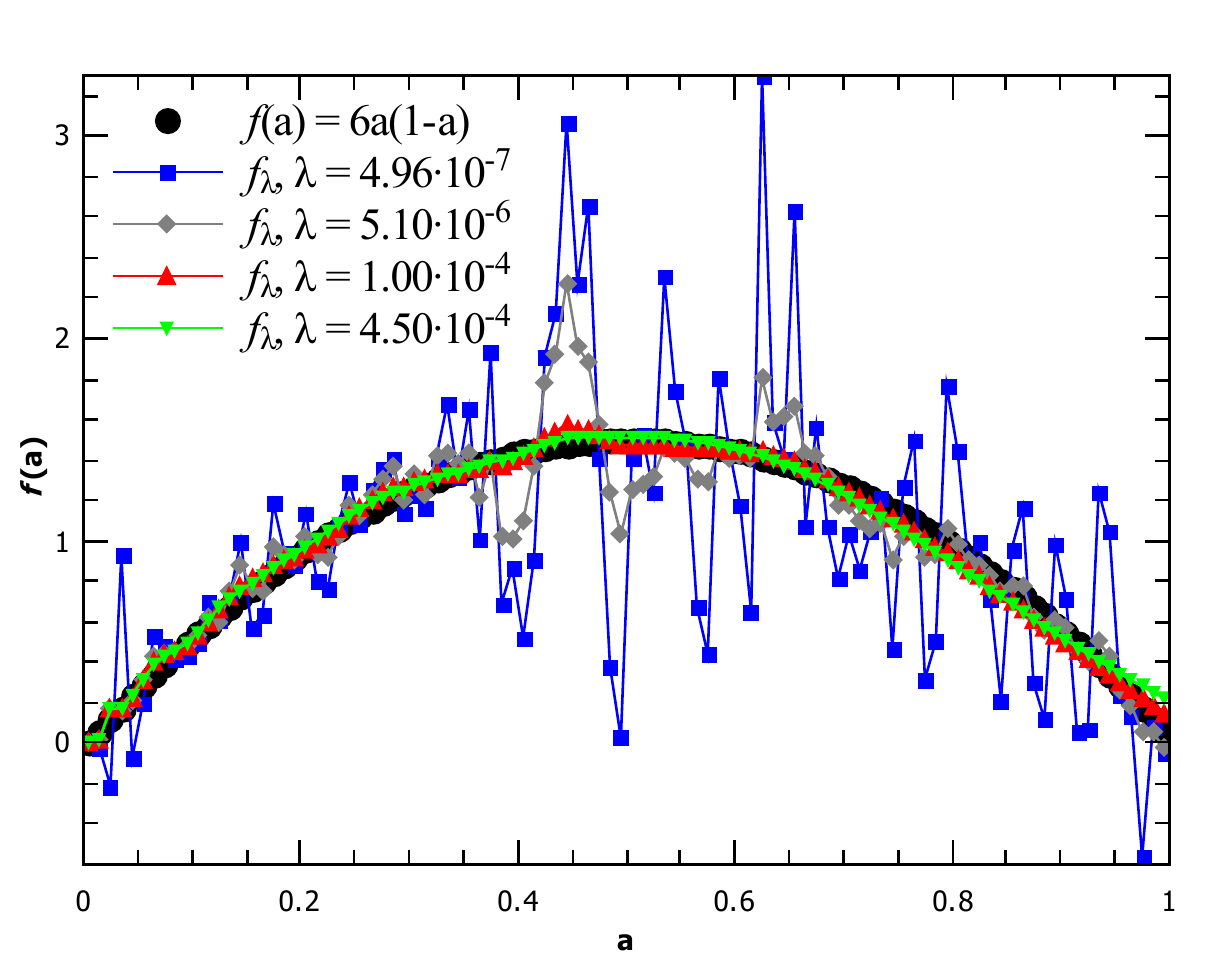}
\centering
\caption{Top: original $\phi$ (black circles) to be inverted: the 
solution of integrating Eqn. \eqref{2} (where $P$ was taken to be 
Eqn. \eqref{3} with $\alpha=1$). After obtaining the Tikhonov 
results of inversion, $f_\lambda$, we integrated them again 
according to Eqn. \eqref{2} to get the shown $\phi_\lambda$; all of 
the curves overlap.
\\
Bottom: Tikhonov results $f_\lambda$ ($\lambda=4.96 \cdot 10^{-7}$, 
$5.1 \cdot 10^{-6}$, $1 \cdot 10^{-4}$ and $4.5 \cdot 10^{-4}$), 
shown with the true $f(a)=6a(1-a)$ (black circles).}
\label{fig:2}
\end{figure}

In Figures \ref{fig:1} and \ref{fig:2} (where $L$ is the identity 
operator), we use $f(a)= 6a(1-a)$ (Equation \eqref{3} with 
$\alpha =1$, corresponding to the deposition case for critical 
island size $i=1$ or the diffusion case for $i=0$) to show the 
results of the Tikhonov regularisation method for a known function. 

With this $f(a)$, we integrated Eqn. \eqref{2} iteratively to obtain 
$\phi(x)$ and solved Eqn. \eqref{1.1} for $f_\lambda$ (solved the 
inverse problem). Figure \ref{fig:1} shows the L-curve, where each 
point of the curve corresponds to a different $\lambda$ ($\lambda$ 
increases from left to the right). Note that 
$\lVert Kf_\lambda -\phi \rVert _2$ is the root mean error between 
the input $\phi$ and $\phi_\lambda$ 
(where $\phi_\lambda =K f_\lambda$ is the result of integrating 
Eqn. \eqref{2} with $f=f_\lambda$).

We chose four values of $\lambda$, highlighted on the L-curve plot, 
and show the four solutions $f_\lambda$ in Figure \ref{fig:2} 
(bottom panel), as well as the corresponding $\phi_\lambda$ (upper 
panel), alongside 
the original $\phi$ and $f$. While different $\phi_\lambda$ lie 
almost perfectly on top of each other and on top of input $\phi$, 
the solutions $f_\lambda$ show how strongly this problem is ill - 
posed.
For the two smaller values of $\lambda$ the solutions $f_\lambda$ 
exhibit high oscillations and the largest $\lambda$ begins to show 
signs of over-smoothing in the interval $a \in [0.7,1]$. The best 
solution still has some noise, it isn't 
symmetric, and needs normalisation; the area under the curve is 
$\lVert f_\lambda \rVert _1 \approx 0.99$. We found that a general 
trend is a decreasing $\lVert \cdot \rVert _1$ norm with growing 
$\lambda$ (moving away from the corner of the L-curve to the right). 
The same problems are amplified when applying the method to 
$\phi_{kMC}$, obtained from (noisy) kMC data, with an additional 
problem that the solutions $f_\lambda$ sometimes dropped slightly 
below zero near $a=1$, although seemingly within the noise error we 
would expect when solving for kMC data input.

Since this implementation of Tikhonov regularisation doesn't give 
entirely satisfactory results (loss of symmetry, 
$\lVert \cdot \rVert _1$ norm or positivity), we would need to 
modify it. Normalization can be always done by hand, but adding 
extra symmetry and positivity constraints on the regularisation, 
while theoretically possible, would turn the L-curve of our 
minimization problem into a 3-dimensional hypersurface 
in ${\mathbb R}^4$. 
This would extend the scope of work enormously so instead we look 
for an alternative approach.

\subsection{Fourier representation for the Inverse Problem}

To complement the Tikhonov regularisation results, we develop an 
alternative method of solving the inverse problem. In this method, 
we represent $f(a)$ as a finite Fourier series whose corresponding 
$\phi$ matches the true, kMC obtained $\phi_{kMC}$.

\begin{figure}[htbp!]
\includegraphics[width=8cm]{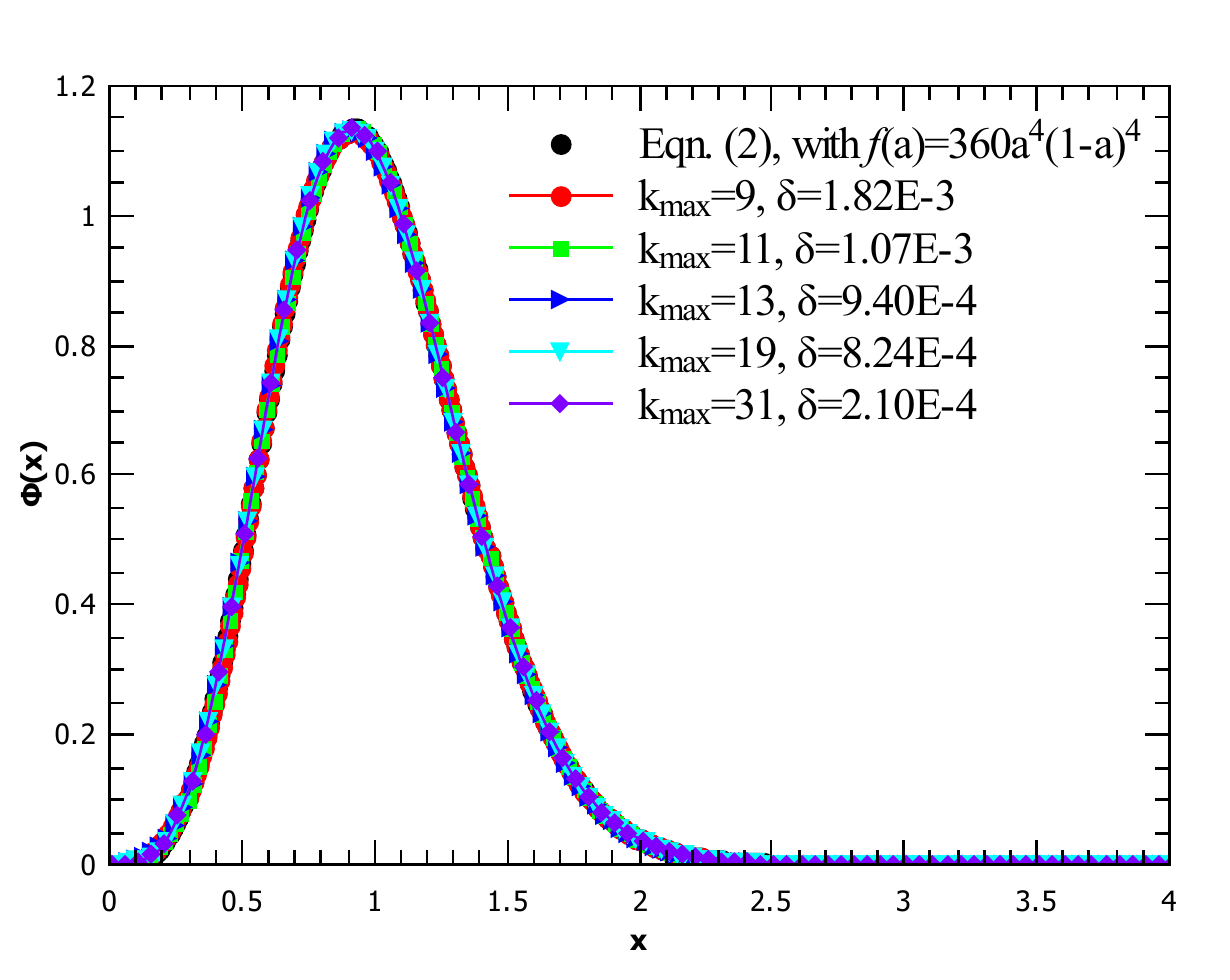}
\centering
\includegraphics[width=8cm]{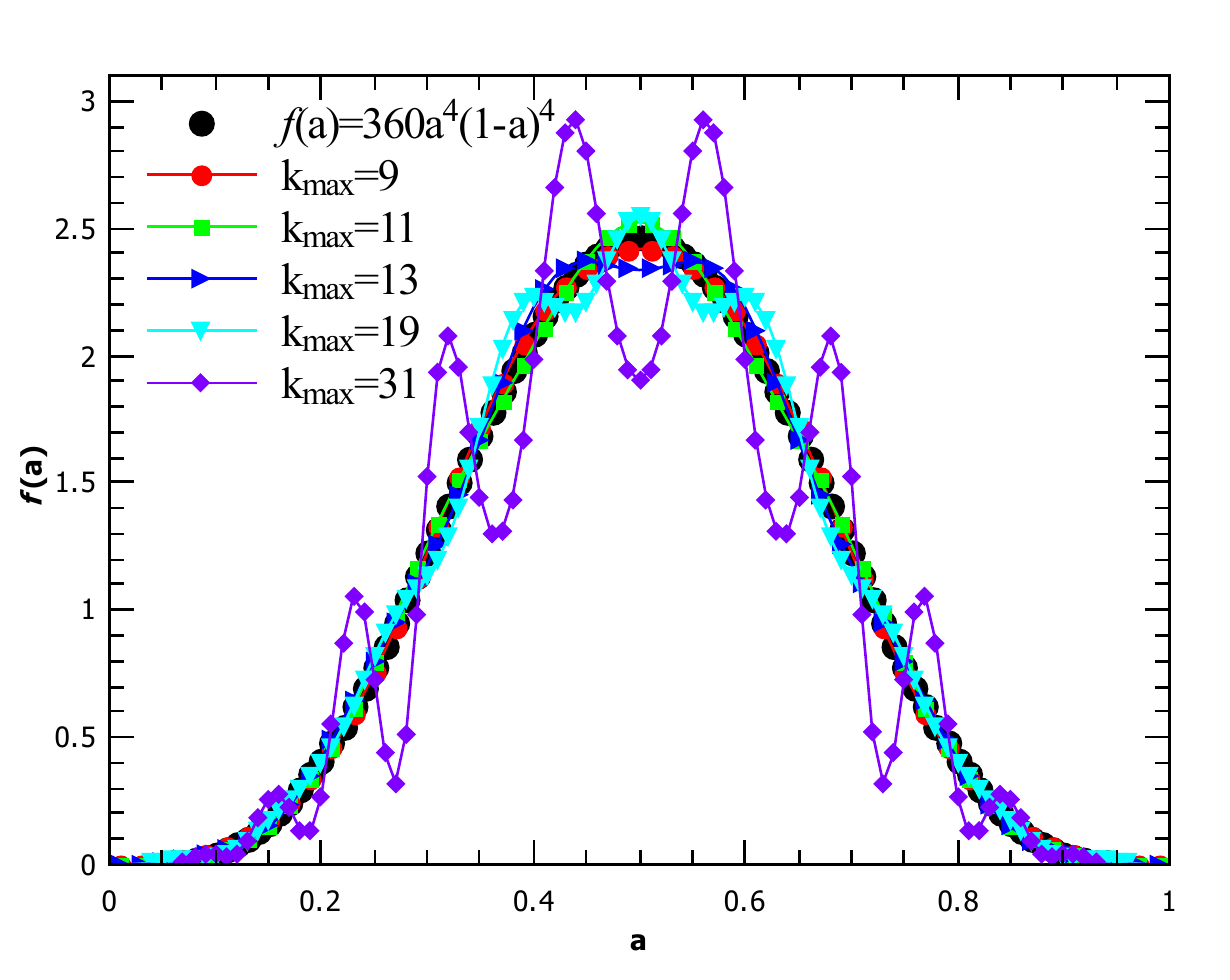}
\centering
\caption{Searching for optimal parameter $k_{max}$: Top: original 
$\phi$ to be inverted (Eqn. \eqref{2} in which $f$ was taken to be 
Eqn. \eqref{3} with $\alpha=4$). After obtaining 
the results of inversion built with Fourier series, $f_F$, for 
different values of $k_{max}$, we integrated them again according to 
Eqn. \eqref{2} to get the $\phi_F$ plotted on top of each other; all 
of the curves overlap.
\\
Bottom: Inversion results $f_F$ ($k_{max}= 9$, $11$, $13$, $19$ and 
$31$), shown with the true $f(a)=360a^4(1-a)^4$.}
\label{fig:2F}
\end{figure}

Whether we take Eqn. \eqref{3} to be an accurate model of physical 
systems or not, $f(a)$ at least has to be equal to zero at $a=0,1$ 
and it is physically reasonable to assume it is symmetrical. 
Therefore we only use sine waves and odd wave numbers to enforce 
symmetry around $a=1/2$ and the requirement $f_F (0)=f_F(1) =0$. We 
start with a single normalized sine; $f_F(a)=N \sin ( \pi a)$, $N$ 
being the normalization constant. We integrate this $f_F$ according 
to Eqn. \eqref{2} to obtain $\phi _F$ and calculate the 
error:
\begin{align} \label{1.2}
\delta = \sum_{i} ( \phi_F (x_i) - \phi_{kMC} (x_i) )^2 .
\end{align}
Then we proceed to build $f_F$ by adding higher random harmonics 
$A \sin (k \pi a)$, where in each step we randomly choose the wave 
number (from allowed values $k = 3,5,...,k_{max}$) and the amplitude 
$A$ ($A \in [-A_{max},A_{max}]$), normalize the new $f_F$ and 
recalculate $\phi_F$ and $\delta$. Then, provided that the resulting 
$f_F$ is everywhere positive, we keep the newly added harmonic with 
the Boltzmann probability $ \exp \left[ - (\delta_{new} - 
\delta_{old}) \beta \right] $. We repeat this cycle with a fixed 
$\beta $ (initially set to 1) $m$ times before increasing $\beta$ by 
a factor of 2 (i.e. perform a simulated anneal). After increasing 
$\beta$ in such a way $M$ times, we narrow in on the solution by a 
search in which we only keep the newly added harmonics if 
$\delta_{new} < \delta_{old}$.

Since there are 5 search parameters (values of $k_{max}$ and 
$A_{max}$, number of cycles $M,m$ and the number of search attempts  
while only accepting moves with $\delta_{new} < \delta_{old}$), we  
needed to find the optimal parameters on a known problem 
before proceeding to calculate $f_F$ for $\phi_{kMC}$. 

Therefore we first integrated Eqn. \eqref{2} with $f(a)$ given by 
Eqn. \eqref{3}, and then used the resulting $\phi$ in place of 
$\phi_{kMC}$ in \eqref{1.2}, to see how can we correctly reconstruct 
$f_F$.
Because Eqn. \eqref{2} is ill posed, adding higher harmonics 
actually leads to a worse, less stable solution $f_F$ with high 
frequency noise, as shown on Figure \ref{fig:2F}. At the same time 
the error $\delta$ (Eqn. \eqref{1.2}) can decrease (here with the 
rest of the search parameters fixed, although in general, when 
increasing $k_{max}$, a higher number of search cycles is needed to 
reach a stable solution). This happens regardless of the amount (or 
absence) of noise in the input and cannot be avoided.
It is a consequence of the following property of the equation 
$Kf=\phi$: the inverse $K^{-1}$ of the operator $K: U \rightarrow V$ 
is unbounded, and the equation is ill posed, if $U$ is an infinite 
dimensional space \cite{4}. Hence decreasing the dimension of space, 
spanned with the harmonics, in which we build $f$, is a form of 
regularisation.

Because of that, we limited the maximum allowed wave number to 11. 
With $k_{max}=11$ and allowed maximum amplitude $A_{max}=0.05$ per 
one search attempt, we ran the simulated anneal with $m=30$ and 
$M=500$ cycles ($30 \times 500$ random harmonic choices) and then 
ran through another 300 attempts, accepting only 
$\delta_{new} < \delta_{old}$. These are the parameters we then 
used to calculate $f_F$ for $\phi_{kMC}$, for all the values of $i$  (for $i=0$ we also use $k_{max}=5$ as explained below).

\section{\label{sec:level14} RESULTS}

\begin{figure}[ht]
\includegraphics[width=8cm]{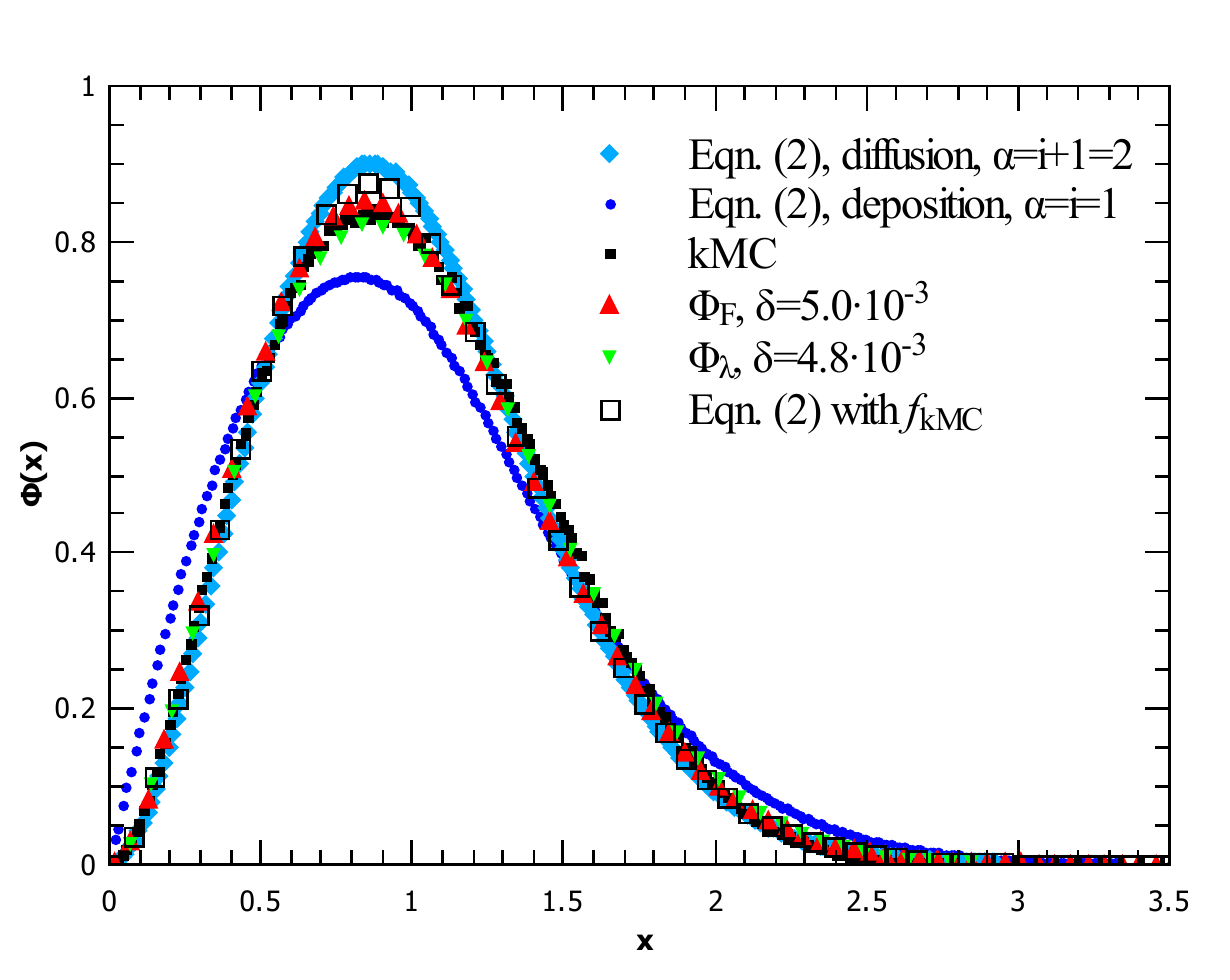}
\centering
\includegraphics[width=8cm]{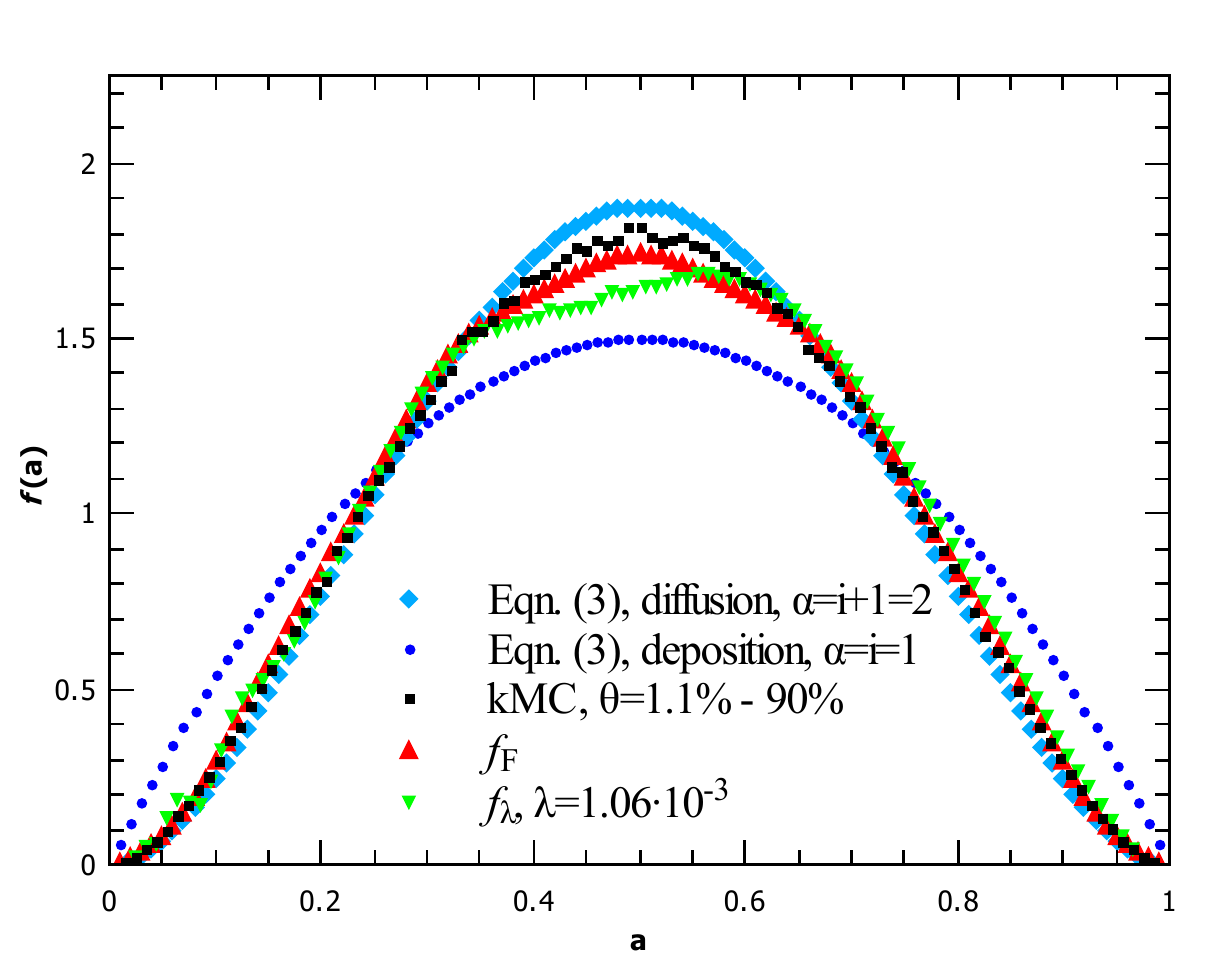}
\centering
\caption{Critical island size $i = 1$:\\
Top: solutions of integrating Eqn. \eqref{2} with $f$ given by 
Eqn. \eqref{3} ($\alpha = i+1$ case in light blue diamonds and $\alpha=i$ in 
dark blue circles). The kMC obtained GSD $\phi_{kMC}$ (full black squares) is inverted 
according to Eqn. \eqref{2}; when the resulting $f_{F, \lambda}$ is 
used to integrate Eqn. \eqref{2} we get $\phi_{F, \lambda}$ (shown 
in red up and green down-facing triangles, respectively). The empty squares show the result 
of integrating Eqn. \eqref{2} with $f = f_{kMC}$. \\
Bottom: $\alpha = i+1$ and $\alpha = i$ case of Eqn. \eqref{3} 
(light blue diamonds and dark blue circles), the kMC result $f_{kMC}$ (full black squares) and the 
results of inverting $\phi_{kMC}$: $f_F$ (red up triangles) and $f_{\lambda}$ 
(green down triangles).}
\label{fig:i1}
\end{figure}

\begin{figure}[ht]
\includegraphics[width=8cm]{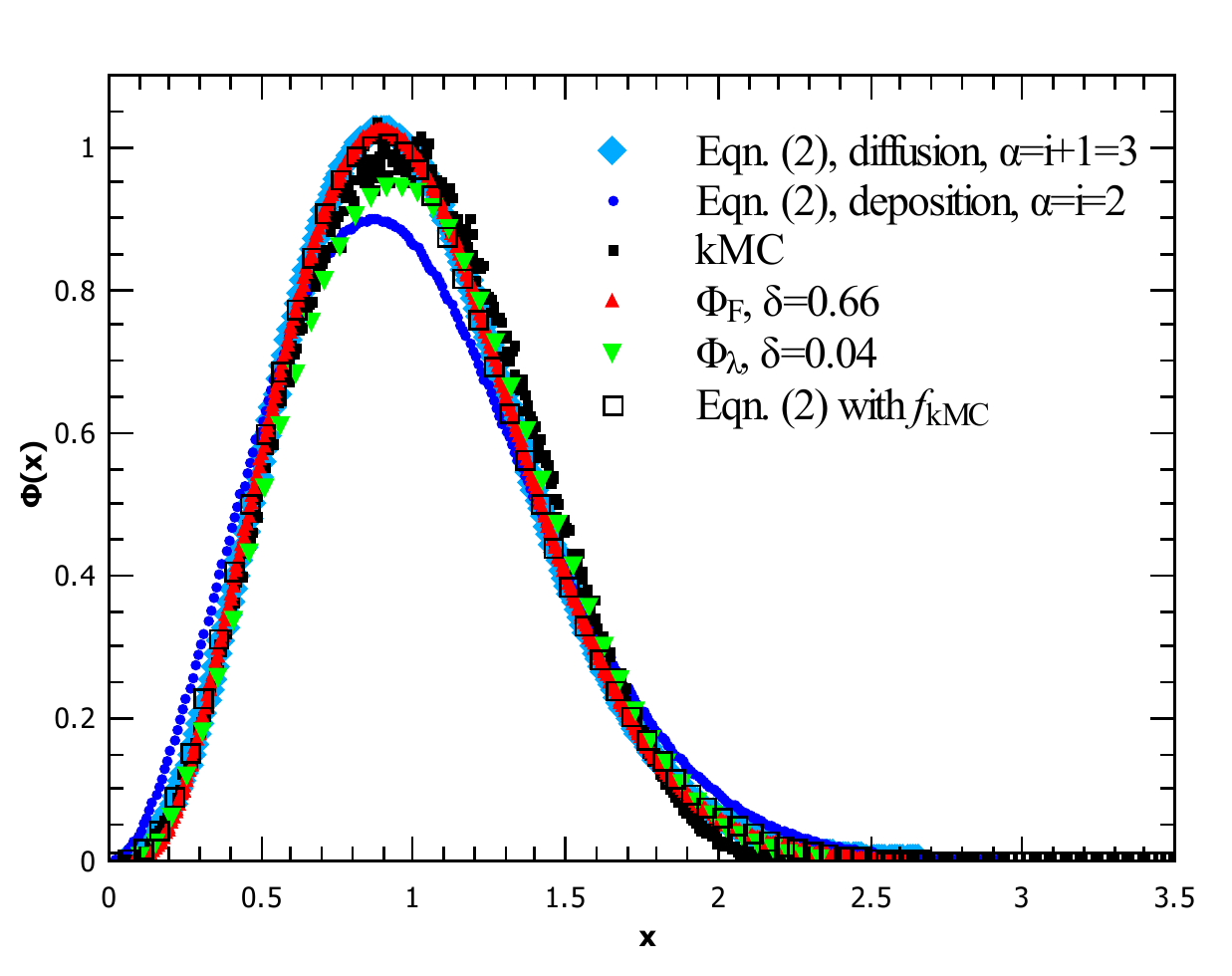}
\centering
\includegraphics[width=8cm]{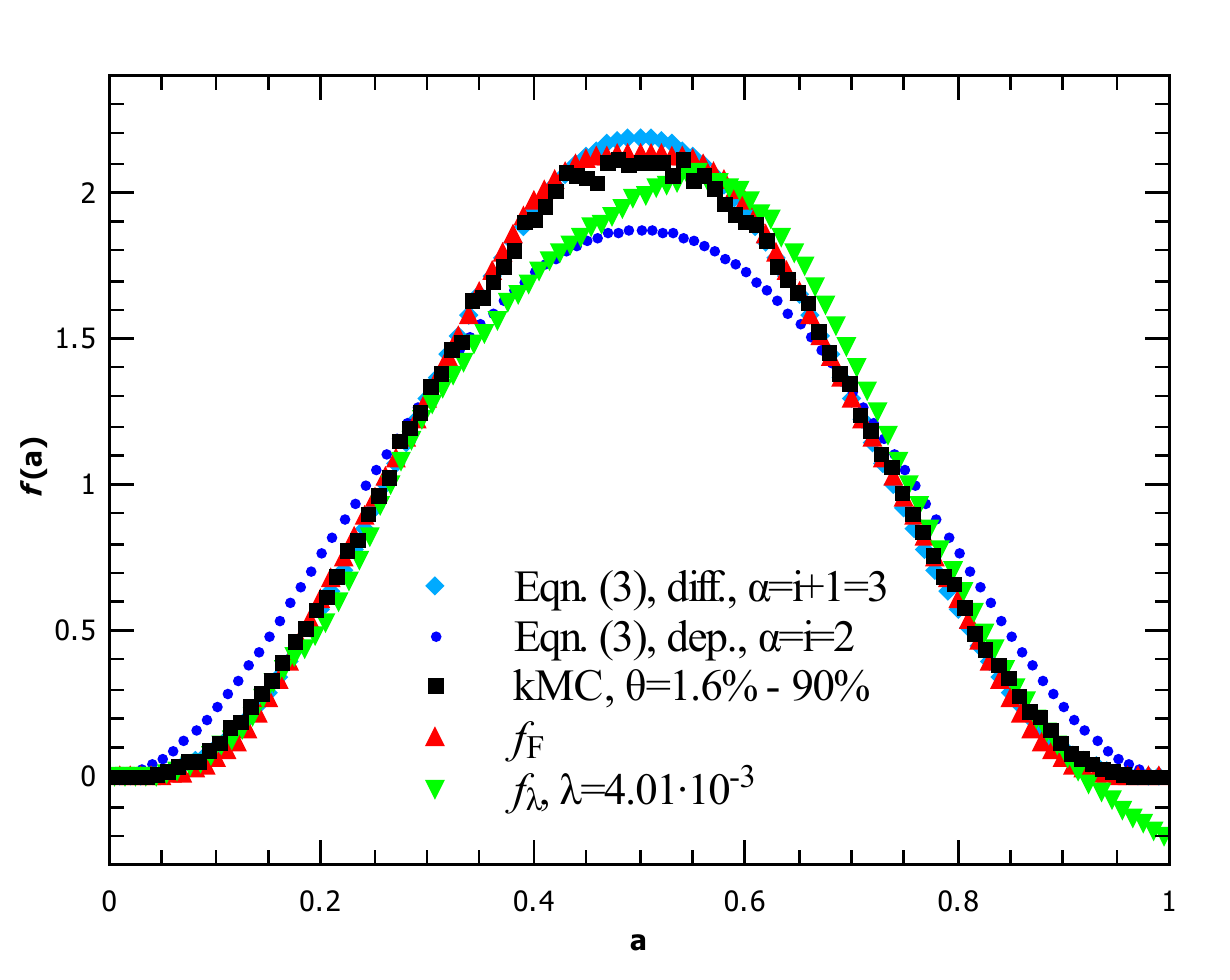}
\centering
\caption{Critical island size $i = 2$. The symbols used in this 
figure have the same meaning as in Fig. \ref{fig:i1}.}
\label{fig:i2}
\end{figure}

\begin{figure}[ht]
\includegraphics[width=8cm]{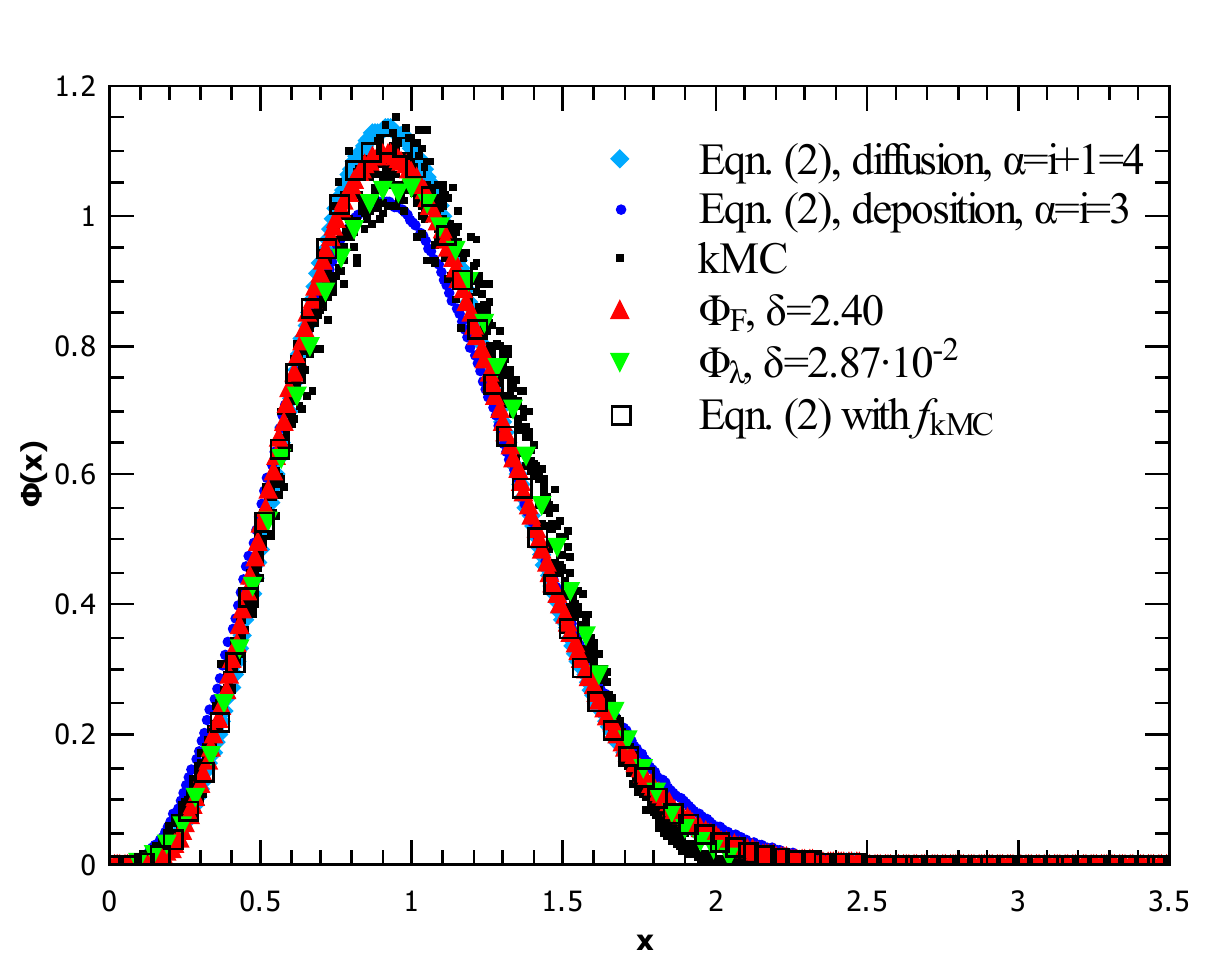}
\centering
\includegraphics[width=8cm]{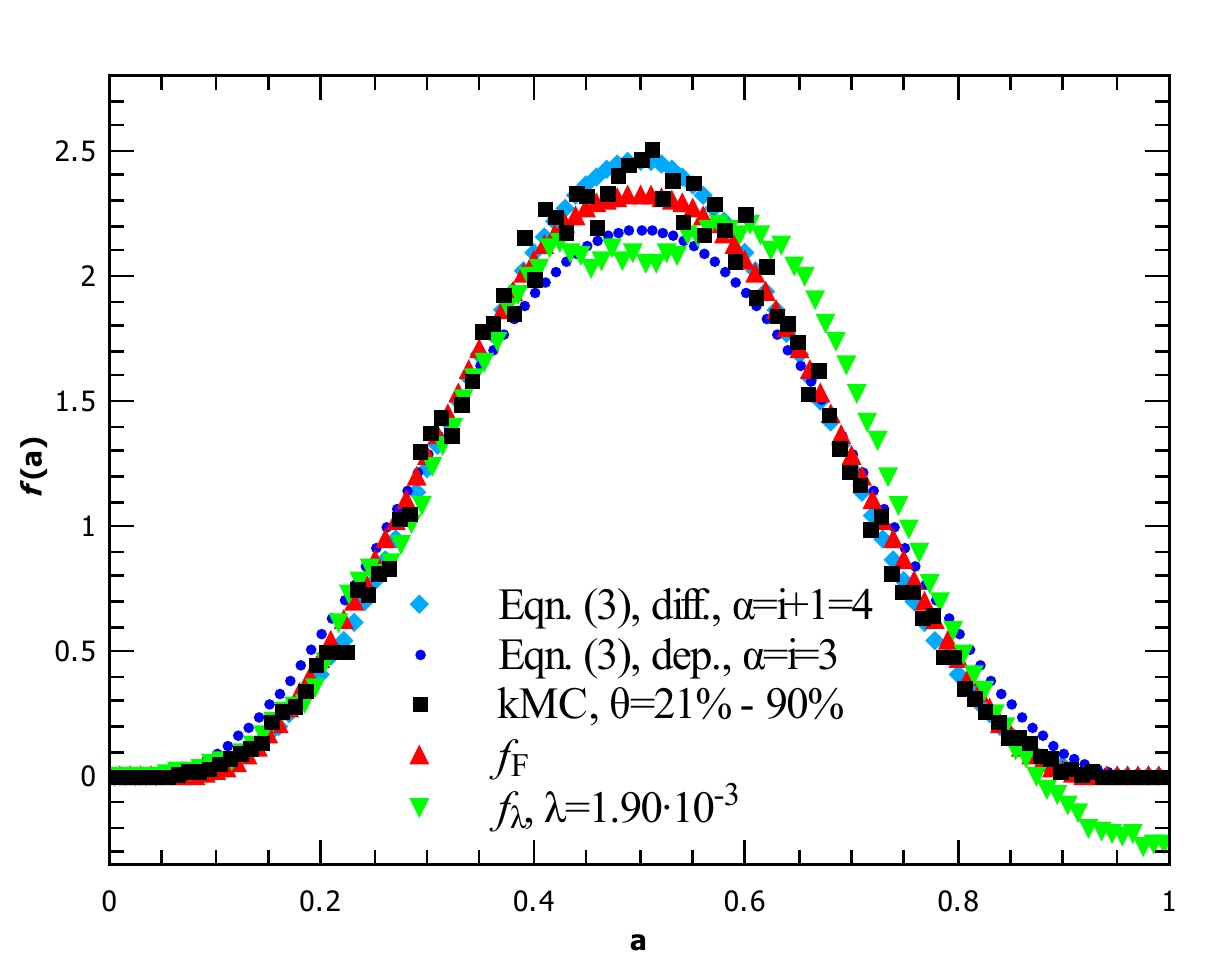}
\centering
\caption{Critical island size $i = 3$. The symbols used in this 
figure have the same meaning as in Fig. \ref{fig:i1}.}
\label{fig:i3}
\end{figure}

\begin{figure}[ht]
\includegraphics[width=8cm]{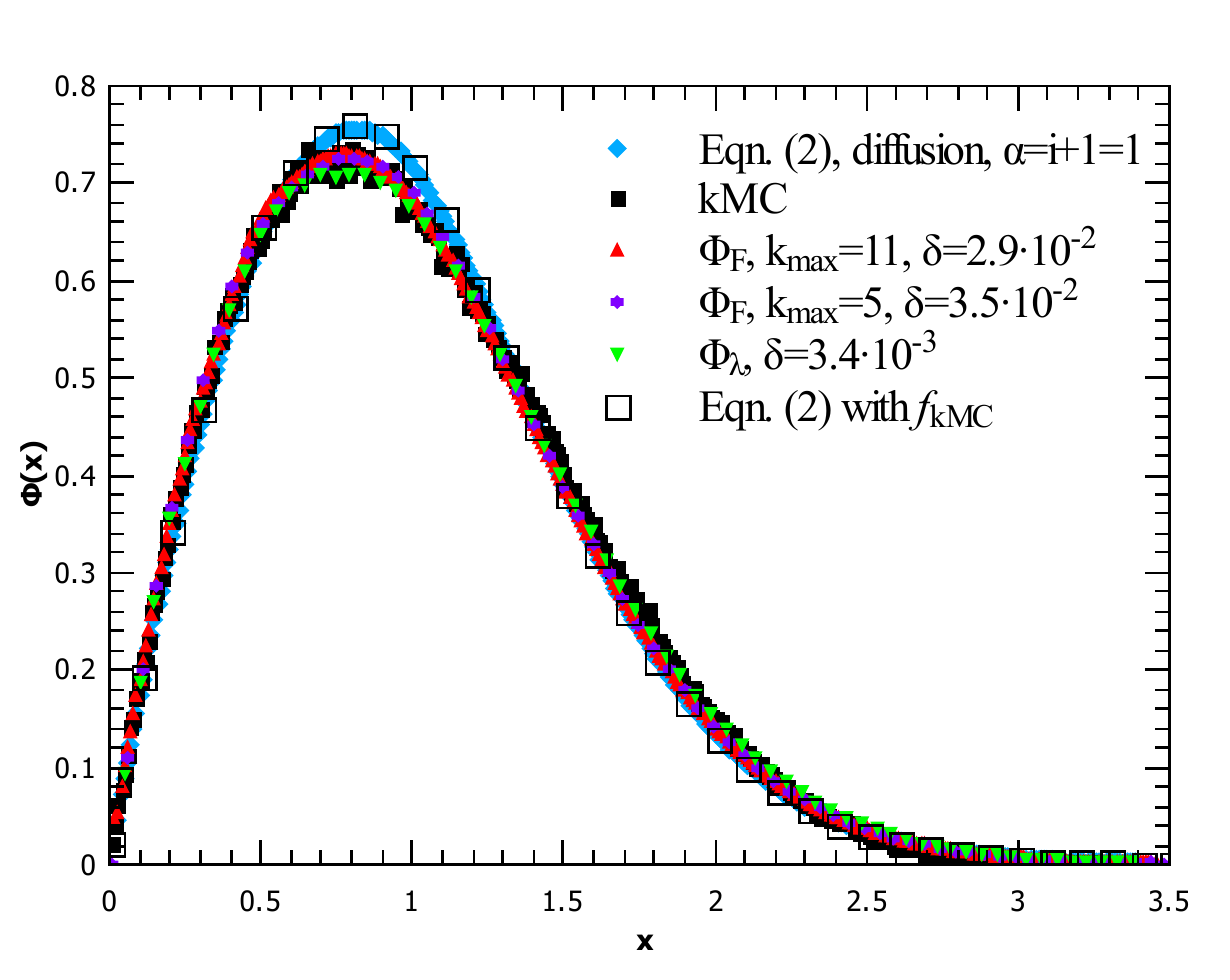}
\centering
\includegraphics[width=8cm]{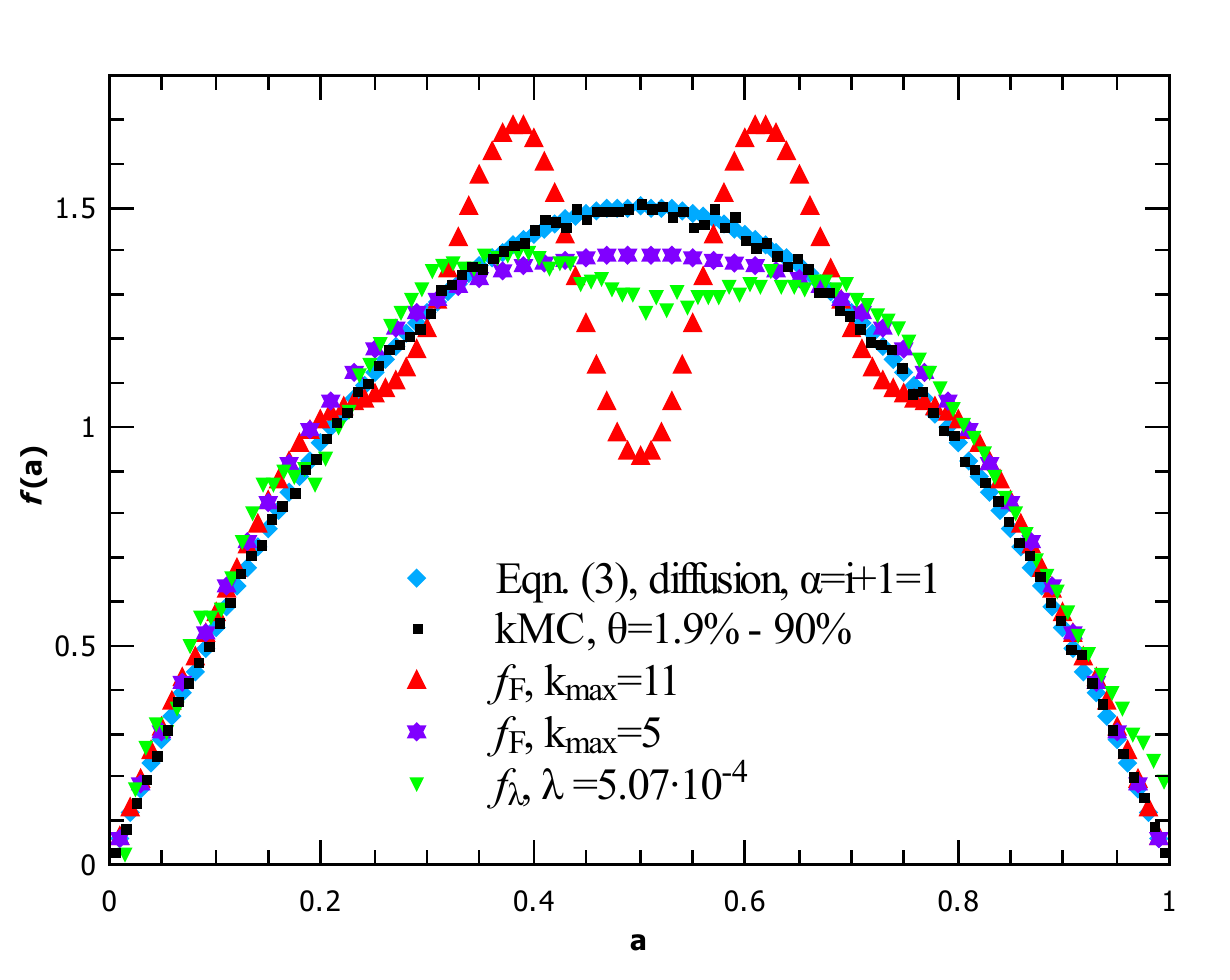}
\centering
\caption{Critical island size $i = 0$. The symbols used in this 
figure have the same meaning as in Fig. \ref{fig:i1}.}
\label{fig:i0}
\end{figure}

We show the diffusion and deposition $\phi$ (Eqn. \eqref{2} with $f$ 
given by two cases of Eqn. \eqref{3}), the GSD  obtained from kMC 
($\phi_{kMC}$), and $\phi_F$, $\phi_\lambda$ plotted together on 
upper panels in Figures \ref{fig:i1},\ref{fig:i2},\ref{fig:i3} 
and \ref{fig:i0}, for critical island size $i=1,2,3$ and $0$ 
respectively. The solutions of integrating Eqn. \eqref{2} with  
$f(a)=f_{kMC}(a)$ are plotted with empty square symbols.

The bottom panels of Figures \ref{fig:i1},\ref{fig:i2},\ref{fig:i3} 
and \ref{fig:i0} show the deposition and diffusion $f(a)$ given by 
Equation \eqref{3}, $f_{kMC}(a)$ obtained from kMC simulations, and 
the solutions of the inverse problem $f_F$ and $f_\lambda$. 

Errors $\delta$ listed in the legends are the sum of squares 
differences between kMC obtained $\phi_{kMC}$ and $\phi_{F}$, 
$\phi_\lambda$ obtained by integrating the solutions $f_{F}$,  
$f_\lambda$ according to Eqn. \eqref{2} (for $\phi_F$ error is given 
with Eqn. \eqref{1.2} and for $\phi_\lambda$ with 
$\lVert Kf_\lambda - \phi_{kMC} \rVert _2 ^2$). The solutions 
$f_F$ are always normalized during the procedure of adding new 
harmonics, but the Tikhonov procedure only deals with the 
$\lVert \cdot \rVert _2$ norm so none of the $f_\lambda$ solutions 
shown have $\lVert \cdot \rVert _1$ norm equal to one. We have 
found, however, that all $i$ solutions with optimal choices of 
$\lambda$ have norm close to 1, and it only significantly drops 
(below 0.95) for too high $\lambda$ which also gave large error 
$\delta$. 

We note here that our $f_{kMC}$ results are similar to the 
nucleation probabilities for $i=1$, $2$ and $3$ shown in a recent 
publication by Gonz\'alez, Pimpinelli and Einstein \cite{EP2017}.

When we use $f_{kMC}$ to integrate Eqn. \eqref{2}, the resulting GSD 
(empty squares in the upper panels of Figures \ref{fig:i1},
\ref{fig:i2},\ref{fig:i3} and \ref{fig:i0}) fits the kMC obtained 
GSD ($\phi_{kMC}$) quite well for all the $i$ cases, but it doesn't 
match it perfectly. We remind the reader here that the DFPE model we 
are using involves a mean field approximation; a non-mean field 
version suggested in Ref. \cite{Ken} gives more accurate results.

Returning to the inverse problem, for the $i=1$ and $2$ cases 
(Figures \ref{fig:i1} and \ref{fig:i2}) both the Fourier and the 
Tikhonov method gave good $f_{F}$, $f_\lambda$ results, but in the 
$i=2$ case we start to see the effect of increased noise in the 
input $\phi_{kMC}$ relative to the $i=1$ case: $f_\lambda$ is 
noticeably negative near $a=1$. 
In the $i=3$ case the situation is even worse (see 
Figure \ref{fig:i3}), so here the Tikhonov solution is more of a 
guideline for the behaviour of the true $f(a)$. On the other hand, 
the Fourier construction was successful in all the cases
so we can conclude that, by using both methods for assurance, we can 
find reliable solutions in problems where $f(a)$ 
is not directly measurable (e.g. many experiments to create 
nanostructures).

In the $i=0$ case (Figure \ref{fig:i0}), only the diffusion limit 
($\alpha=i+1$) of Eqn. \eqref{3}, as introduced in Ref. \cite{BM}, has 
physical meaning. In addition, the Fourier result for $\phi_{kMC}$ 
inversion with $k_{max}=11$ is problematic. Its high oscillations 
around $a=0.5$ suggest a higher degree of regularisation is needed, 
so even though the previously established cut-off $k_{max}=11$ gave 
excellent results when inverting Eqn. \eqref{2} for all 
$\alpha$ values in Eqn. \eqref{3} (including the here relevant 
$\alpha=i+1=1$), we additionally show the inverse $f_F$ where we 
used $k_{max}=5$. This result is backed by the Tikhonov solution ($\lambda$ is taken from the corner area of the L-curve).

The measured $f_{kMC}(a)$ for $i=0$ lies almost perfectly on top of 
the diffusion curve. 
However, the solution of Eqn. \eqref{2} with $f=f_{kMC}$ is (in this 
case most noticeably) not matching $\phi_{kMC}$, which shows the 
limitations of the mean field approximation used to formulate this 
approach.
 
For critical island sizes higher than 0, the measured $f_{kMC}(a)$ 
(and, consequently, $\phi_{kMC}$) is at least a little below the 
diffusion prediction, allowing for a small contribution of the 
deposition driven nucleation. 
We finish our analysis by quantifying the level of this contribution 
for different $i$.
Table \ref{table:1} shows the result of fitting $f_{kMC}$ on a 
convex combination of analytic expressions for diffusion and 
deposition $f(a)$ from Eqn. \eqref{3}:
\begin{align} \label{linkomb}
f_{kMC} = \beta f_{\alpha=i+1} ^{diffusion}  + (1- \beta) 
f_{\alpha=i} ^{deposition}
\end{align}
with the least squares method.
We also show the fit of $\phi_{kMC}$ on the convex combination of 
the diffusion and deposition case,
\begin{align}\label{linkomb2}
\phi_{kMC} = \gamma \phi_{\alpha = i+1} ^{diffusion} + (1- \gamma) 
\phi_{\alpha=i} ^{deposition},
\end{align}
where we fitted kMC curves $\phi_{kMC}$ on the results of numerical 
integration of Eqn. \eqref{2}.
From the results, we can safely conclude that diffusion is the 
dominant mechanism of island nucleation. 
(Note that the result for $\gamma$ in the $i=2$ case is larger than 
1, but not if its allowed error is subtracted.)

\begin{table}[htbp!] 
\begin{center}
\begin{tabular}{ |c|c|c| } 

\hline
$ i $ & $\beta $ & $\gamma$   \\
\hline

1	& 0.821 $\pm$ 0.007	& 0.728 $\pm$ 0.010 \\
2	& 0.819 $\pm$ 0.015	& 1.015 $\pm$ 0.018	   \\
3	& 0.844 $\pm$ 0.004	& 0.714 $\pm$ 0.020	 \\

\hline
\end{tabular}
\end{center}
\caption{Results of fitting kMC results according to Eqns. \eqref{linkomb} and \eqref{linkomb2}.}
\label{table:1}
\end{table}

\section{\label{sec:level15} SUMMARY}

In this paper, we have revisited the mean field DFPE \eqref{1} 
model of gap fragmentation on a one dimensional substrate from 
Ref. \cite{Ken}. Using the Tikhonov regularisation method, from the 
kMC obtained GSD and the integral equation form of the DFPE for the 
GSD (Eqn. \eqref{2}) we were able to calculate the gap fragmentation 
probability; that is, the probability of a new island nucleation 
occurring at a position $a$ inside a gap ($f(a)$). The results show 
fair agreement with the probability  $f_{kMC}(a)$ that  we measured 
directly from kMC simulations, although they lack the expected 
symmetry and strict positivity. Growing amounts of numerical noise 
(in cases of higher $i$) aggravate this problem.

We developed an alternative method of inverting Eqn. \eqref{2} to 
obtain $f$, in which we represent $f$ as a finite Fourier series and 
use the series properties. This allows us to impose symmetry and 
positivity, however a downside is a more time consuming procedure 
due to the large amount of search parameters. The results of this 
method are in better agreement with the measured $f_{kMC}(a)$ so 
this method, especially when backed by the well - known Tikhonov 
method, makes for a good tool in solving problems where it is not 
possible to measure $f(a)$ directly. 

The DFPE model we use involves two limiting cases of island 
nucleation: diffusion (via colliding adatoms) and deposition driven. 
As expected, within this framework our results (both the kMC 
obtained GSD and $f_{kMC}$) favour the 
diffusion driven nucleation as the dominant mechanism. We found no 
correlation between $i$ and one mechanism's contribution amount 
relative to the other, however if there were a trend, a model with a 
built in mean field approximation would most likely be too crude for 
it to be observed, especially from noisy data. 

Finally, we emphasize that, while the DFPE we employ here may not 
offer a perfect fit (as seen with the solutions of Eqn. \eqref{2} 
with $f_{kMC}$ which have slightly higher peaks than $\phi_{kMC}$), 
its strength lies in the unique possibility of calculating $f(a)$ 
from a given GSD, without the need for additional information.



\input{PaperArXiv.bbl}

\end{document}

%% file: PaperArXiv.bbl
%